\documentstyle[prl,aps,epsf,multicol]{revtex}
\begin{document}
\title{\bf Persistence of Manifolds in Nonequilibrium Critical Dynamics}
\author{Satya N. Majumdar$^1$ and Alan J. Bray$^{1,2}$}
\address{$^1$Laboratoire de Physique Quantique (UMR C5626 du CNRS), 
Universit'e Paul Sabatier, 31062 Toulouse Cedex, France \\
$^2$Department of Physics and Astronomy, University of 
Manchester, Manchester M13 9PL, UK}

\date{\today}

\maketitle

\begin{abstract}
We  study  the persistence  $P(t)$  of  the  magnetization of  a  $d'$
dimensional  manifold   (i.e.\  the  probability   that  the  manifold
magnetization does  not flip  up to time  $t$, starting from  a random
initial condition)  in a $d$-dimensional  spin system at  its critical
point. We  show analytically that  there are three distinct  late time
decay forms for $P(t)$  : exponential, stretched exponential and power
law,  depending  on  a  single  parameter  $\zeta=(D-2+\eta)/z$  where
$D=d-d'$  and   $\eta$,  $z$  are  standard   critical  exponents.  In
particular,  our  theory  predicts  that  the persistence  of  a  line
magnetization decays  as a power law  in the $d=2$ Ising  model at its
critical point. For the $d=3$ critical Ising model, the persistence of
the plane  magnetization decays as a  power law, while that  of a line
magnetization  decays as a  stretched exponential.   Numerical results
are consistent with these analytical predictions. 

\noindent

\medskip\noindent {PACS numbers: 05.70.Ln, 05.50.+q, 05.70.Jk}
\end{abstract}

\begin{multicols}{2}

Following a rapid quench of a spin system from infinite temperature to
zero  temperature, domains of  competing ground  states form  and grow
with time\cite{Bray}.  In the  standard nonconserved dynamics  a given
spin inside  the sample flips only  when a domain  wall passes through
it, which happens  rather rarely. As a result,  persistence, i.e.\ the
probability  that the  spin  remains unflipped  up to  time $t$  decays
slowly   as   a   power    law,   $P(t)\sim   t^{-\theta}$   at   late
times\cite{Review}.  In  contrast, if the  spin system is  quenched to
its critical temperature $T_c$,  any given spin fluctuates rapidly due
to the finite temperature and the  persistence of a single spin has an
exponential tail  at late times. However,  for a quench  to $T_c$, the
persistence of  the total magnetization  of the sample (as  opposed to
that  of  a  single spin)  again  decays  as  a power  law,  $P(t)\sim
t^{-\theta_g}$, where the exponent $\theta_g$  has been argued to be a
new   nonequilibrium    critical   exponent\cite{MBCS}.   The   global
persistence   has  since   been   studied  in   a   wide  variety   of
systems\cite{globalp}.

In  a  $d$-dimensional sample,  a  single  spin  is a  $0$-dimensional
manifold.  On the other  hand, the global magnetization corresponds to
summing  over   all  the  spins  of   the  full  sample   which  is  a
$d$-dimensional manifold. To interpolate  between these two limits, it
is  natural  to  study  the  persistence of  the  magnetization  of  a
$d'$-dimensional manifold with $0 \leq  d'\leq d$.  In the  two limits
$d'=0$  and  $d'=d$, the  persistence  has  very different  asymptotic
decay, respectively exponential and power law.  The following question
then naturally arises:  if one tunes the manifold  dimension $d'$ from
$d'=0$  to $d'=d$,  how does  the asymptotic  behavior  of persistence
changes from  an exponential ($d'=0$)  to a power law  decay ($d'=d$)?
Does this change occur abruptly at some intermediate value of $d'$, or
is there an intermediate regime  of $d'$ where the behavior is neither
exponential nor power law, but something in between? In this Letter we
address this  interesting issue and  we show analytically  that indeed
there is a novel intermediate regime of $d'$ where the persistence has
a stretched  exponential tail. Our  main results can be  summarized in
terms of a single number  $\zeta= (D-2+\eta)/z$, where $D=d-d'$ is the
co-dimension  of the  manifold and  $\eta$  and $z$  are the  standard
critical exponents:  $z$ is the dynamical exponent  that describes the
temporal  growth of  the correlation  length $\xi(t)\sim  t^{1/z}$ and
$\eta$  describes the  power law  decay of  the  equilibrium spin-spin
correlation  function at $T_c$,  $\langle \phi  (0)\phi(r)\rangle \sim
r^{-(d-2+\eta)}$ for  large separation $r$. Depending on  the value of
$\zeta$, the  persistence of  the magnetization of  a $d'$-dimensional
manifold at $T_c$ has the asymptotic behavior, 
\begin{eqnarray}
P(t)   &\sim  &   t^{-\theta(d',d)},   \quad\quad\quad\quad\,  \zeta<0
     \nonumber  \\ &\sim  & \exp\left(-a_1  t^{\zeta}\right), \,\,\,\,
     0\leq \zeta\leq 1 \nonumber  \\ & \sim & \exp\left(-b_1 t\right),
     \quad\quad\quad \zeta>1 ,
\label{asymp1}
\end{eqnarray}
where the exponent  $\theta(d',d)$ depends on $d'$ and  $d$ and $a_1$,
$b_1$ are  constants.  Strictly  speaking, in the  intermediate regime
$0\leq \zeta\leq  1$ we will show  that, $\exp\left(-a_2 t^{\zeta}\ln
t\right)\leq  P(t)  \leq  \exp\left(-a_1 t^{\zeta}\right)$  for  large
$t$.  We will  derive the  results  in Eq.\ (\ref{asymp1}) within  the
mean-field theory (valid for $d>4$), in the $n\to \infty$ limit of the
$O(n)$ model, followed by a general scaling theory.

Two specific applications  of our general results are  as follows: (i)
Consider  the persistence of  the line  magnetization ($d'=1$)  in the
$d=2$  Ising  model at  $T_c$.  Using  $d'=1$,  $d=2$, $\eta=1/4$  and
$z\approx  2.172$, one gets  $\zeta=(d-d'-2+\eta)/z\approx -0.3453<0$.
Hence  Eq.\  (\ref{asymp1})  predicts   a  power  law  decay  for  the
persistence of the line magnetization; (ii) For the $d=3$ Ising model,
using $\eta\approx 0.032$  and $z\approx 2$, one finds  that while for
the plane  magnetization ($d'=2$),  $\zeta\approx -0.484 <0$,  for the
line  magnetization  ($d'=1$),  $0<\zeta\approx  0.016<1$.   Our  Eq.\
(\ref{asymp1}) then predicts a power  law decay for the persistence of
the plane magnetization, but a stretched exponential decay for the line
magnetization.  Numerical simulations for small samples in $2$ and $3$
dimensions are consistent with these analytical predictions.

Our starting  point is the  standard Langevin equation for  the vector
order parameter $\vec \phi=(\phi_1,\ldots, \phi_n)$,
\begin{equation}
\partial_t\phi_i=\nabla^2  \phi_i  -r\phi_i -(u/n)\phi_i  \sum_{j=1}^n
\phi_j^2 +\eta_i,
\label{Lange1}
\end{equation}
where ${\vec \eta}  ({\bf x}, t)$ is a Gaussian  white noise with mean 
zero and correlator  $\langle   \eta_i({\bf   x},t)\eta_j({\bf
x'},t')\rangle=2\delta_{i,j}\delta^d({\bf                      x}-{\bf
x'})\delta(t-t')$.  The magnetization  of a  $d'$-dimensional manifold
will be defined by  the vector field, $\psi_i (x_{d'+1},\ldots, x_d,t)
=\int  \phi_i(\vec x  ,t)\prod_{i=1}^{d'}{ {dx_i}\over  {\sqrt  L} }$,
obtained   by  integrating   the   order  parameter   over  the   $d'$
directions.  Here $L$  denotes  the  length of  the  sample along  any
direction.  For  a vector order  parameter, we define  the persistence
$P(t)$ of the manifold to  be the probability that any given component
of the manifold magnetization, say $\psi_1$, does not change sign up 
to time $t$. Since  all the  components of  the  spin are  equivalent,
henceforth we will drop the subscript $i$ of $\psi_i$ for convenience.
Note  that the  magnetization $\psi$  is  a field  over the  remaining
$D=d-d'$ dimensional space  whose coordinates $(x_{d'+1},\ldots, x_d)$
will be  relabelled by the  vector $\vec r= (r_1,r_2,\ldots  r_D)$ for
convenience.  The  observation  that  allows  us  to  make  analytical
predictions for the persistence  of the magnetization $\psi(\vec r,t)$
is that  it is a Gaussian  variable at all finite  times. This follows
simply  from the  fact that  $\psi(\vec r,  t)$ is  a sum  of $L^{d'}$
random variables  which are  correlated but only  over a  {\em finite}
correlation  length $\xi(t)\sim  t^{1/z}$. Thus  in  the thermodynamic
limit  when $t^{1/z}\ll L$,  the central  limit  theorem asserts  that
$\psi(\vec r  ,t)$ is a Gaussian  field. Hence its persistence is 
determined by the  autocorrelation function  $C(t_1,t_2)=\langle 
\psi(\vec r,t_1)\psi(\vec r,t_2)\rangle$\cite{Review}.

We start with the mean-field theory, valid for $d\geq 4$, where we set
$u=0$,    and    also   $r=0$    (at    the    critical   point)    in
Eq.\ (\ref{Lange1}).  Next we integrate the Langevin  equation over the
$d'$ space directions and then  solve the resulting linear equation in
the Fourier  space. Defining ${\tilde \psi}(\vec  k, t)=\int \psi(\vec
r, t)e^{i {\vec k}\cdot{\vec r}}d^Dr$, it is easy to compute the 
two-time correlation function,
\begin{eqnarray}
\langle {\tilde \psi}(\vec k,  t_1) {\tilde \psi}(-\vec k, t_2)\rangle
&=&  \Delta   (\vec  k)  e^{-k^2(t_1+t_2)}\nonumber   \\  &+&  {1\over
{k^2}}\left(e^{-k^2|t_1-t_2|}-e^{-k^2(t_1+t_2)}\right),
\label{corrk1}
\end{eqnarray}
where  $\Delta(\vec  k)=\langle   {\tilde  \psi}(\vec  k,  0)  {\tilde
\psi}(-\vec k, 0)\rangle$ is taken  to be a constant (for uncorrelated
initial  condition). At  late times,  the initial condition dependent
term  becomes negligible  and it  is sufficient to retain  only the
second  term  on  the  right-hand side  of  Eq.\  (\ref{corrk1}).  The
autocorrelation  function is  then  obtained by  integrating over  the
$\vec  k$  space, $C(t_1,t_2)=\int  d^Dk\,e^{-k^2a^2} \langle  {\tilde
\psi}(\vec k,  t_1) {\tilde \psi}(-\vec k, t_2)\rangle$  where we have
introduced a soft ultraviolet cut-off $a$. One finds
\begin{equation}
C(t_1,t_2) = \beta A\left[(t_1+t_2+a^2)^{2\beta}
-(|t_1-t_2|+a^2)^{2\beta}\right],
\label{corr2}
\end{equation} 
where $2\beta=(2-D)/2$ and $A$ is an unimportant constant.

Consider first the case $D<2$,  i.e., $\beta>0$. In this case there is
no  need for  the ultraviolet  cut-off since  $\langle  \psi^2(\vec r,
t)\rangle =C(t,t)$  does not  diverge at any  finite time even  if one
puts  $a=0$. Setting  $a=0$  in  Eq.\ (\ref{corr2})  we  find that  the
correlation  function $C(t_1,t_2)$  is still  non-stationary. However,
one   can   render  it   stationary   by   employing   a  well   known
trick\cite{Review},  where   one  introduces  a   normalized  Gaussian
process, $X=\psi/\sqrt{\langle \psi^2\rangle}$ which, when observed in
the logarithmic time $T=\ln  t$ becomes a stationary Gaussian variable
with    the   correlator,   $C(T)=\langle    X(0)X(T)\rangle   ={\cosh
(T/2)}^{2\beta}-|\sinh(T/2)|^{2\beta}$.  Interestingly,   the  same
Gaussian  correlator also  appears in the  context  of the persistence 
of the Edwards-Wilkinson type rough interfaces\cite{interface}. 
It is well known\cite{Slepian,Review} that
for such a stationary  Gaussian correlator (decaying exponentially for
large $T$),  the persistence of the process  also decays exponentially
for large $T$, $P(T)\sim \exp(-\theta  T)$.  In terms of the real time
$t=e^T$, this indicates a power law decay, $P(t)\sim t^{-\theta}$, 
with persistence   exponent  $\theta$.   For   this  particular
correlator, the exponent $\theta$ has  been studied in great detail in
the context of the  interface problem\cite{interface} and the exponent
is   known  to   depend  continuously   on  the   roughness  parameter
$\beta=(2-D)/4 >0$.

For $D>2$,  on the other hand,  one needs the  ultraviolet cut-off $a$
explicitly in order to  keep $\langle \psi^2(\vec r, t)\rangle=C(t,t)$
finite. In  that case, the  appropriate scaling limit is  $t_1, t_2\to
\infty$,  but  keeping their  difference  $|t_1-t_2|$  fixed. In  this
limit, Eq.\ (\ref{corr2}) reduces  to a  stationary correlator  in the
original  time variable, $  C(t_1,t_2)\sim (|t_1-t_2|+a^2)^{-(D-2)/2}$
that decays  as a  power law for  large $|t_1-t_2|$. To  calculate the
persistence  of Gaussian  stationary processes  with  an algebraically
decaying correlator  is nontrivial.  However, there  exists a powerful
theorem due  to Newell and  Rosenblatt\cite{NR}, which states  that if
the  stationary  correlator  decays  as  $C(t)\sim  t^{-\alpha}$  with
$\alpha>0$   for  large  time   difference  $t=|t_1-t_2|$,   then  the
persistence $P(t)$ (probability of  no zero crossing between $t_1$ and
$t_2$) of such  a process has the following  asymptotic behaviors: (i)
$P(t)\sim   \exp(-K_1   t)$   if   $\alpha>1$  and   (ii)   $\exp(-K_2
t^{\alpha}\ln t)  \leq P(t)\leq  \exp(-K_3 t^{\alpha})$ if  $0< \alpha
<1$, where the $K_i$'s are constants.  Applying this  theorem  to  our
problem,  we  find  that  $P(t)\sim  \exp(-K_1 t)$  for  $D>4$  and  $
\exp(-K_2 t^{(D-2)/2}\ln t)  \leq P(t)\leq \exp(-K_3 t^{(D-2)/2})$ for
$2< D <4$.  In the borderline case $D=4$, there  will be an additional
logarithmic correction.

Combining these results for $D<2$  and $D>2$ and noting that $z=2$ and
$\eta=0$ within the mean-field theory, we find that the explicit exact
results for the  mean-field theory derived above are  just the special
cases of  the general result  in Eq.\ (\ref{asymp1}) provided one uses
the mean-field value $\zeta=(D-2)/2$ in Eq.\ (\ref{asymp1}).

The mean-field theory  is valid for $d\geq 4$. In  order to access the
physically  relevant dimensions  $d\leq  4$, we  now consider  another
solvable   limit,   namely  the   $n\to   \infty$   limit  where Eq.\ 
(\ref{Lange1}) becomes
\begin{equation}
\partial_t\phi_i=\nabla^2 \phi_i -[r+S(t)]\phi_i +\eta_i,
\label{Lange2}
\end{equation}
where   $S(t)=u\langle   \phi_i^2\rangle$   has   to   be   determined
self-consistently.     The    critical     point     corresponds    to
$r+S(\infty)=0$. This  self-consistent determination of  $S(t)$ can be
done using  standard techniques\cite{MBCS} and one finds  that at late
times, $S(t)\to S(\infty)- (4-d)/{4t}$ for $2< d \leq 4$. Substituting
this result into Eq.\ (\ref{Lange2}),  summing  over  the  $d'$
directions, solving  the resulting equation  in the Fourier  space and
finally  integrating over  the $\vec  k$ space  (as in  the mean-field
theory), we  finally arrive at the  following autocorrelation function
for the manifold magnetization $\psi(\vec r,t)$ in $2< d \leq 4$,
\begin{equation}
C(t_1,t_2)=  A_1  (t_1t_2)^{(4-d)/4}\int_0^{t_1} {{dt'\,{t'}^{(d-4)/2}
}\over {(t_1+t_2-2t'+a^2)^{D/2}} },
\label{auto1}
\end{equation}
where $t_1\leq t_2$,  $A_1$ is a constant and  $a$ represents the soft
ultraviolet cut-off as before.
 
For $D<2$, as in the mean-field  theory, one can set the cut-off $a=0$
and the resulting nonstationary correlator in Eq.\ (\ref{auto1}) can be
reduced  to  a  stationary   correlator  for  the  normalized  process
$X=\psi/\sqrt{\langle\psi^2\rangle}$ in the logarithmic time $T=\ln t$,
\begin{equation}
A(T)=\left[\cosh(T/2)\right]^{\mu-D/2}
{{B\left[\mu,2\beta,2/(1+e^T)\right]}\over{B\left[\mu,2\beta\right]}},
\label{auto2}
\end{equation}
where $\mu=(d-2)/2$,  $2\beta=(2-D)/2$, $B[m,n]$ is  the standard Beta
function  and  $B[m,n,x]=\int_0^x  dy y^{m-1}(1-y)^{n-1}$.  Since  the
stationary  correlator in Eq.\ (\ref{auto2}) decays  exponentially for
large $T$, $A(T)\sim \exp[-(d+D-2)T/4]$, one concludes 
\cite{Slepian,Review} that the corresponding persistence also
decays  exponentially for  large $T$,  $P(T)\sim \exp(-\theta  T)$ and
hence as a  power law  in  the original  $t=e^T$ variable,  $P(t)\sim
t^{-\theta}$. Determining the  exponent $\theta$ analytically is still
a  challenging task. However  one can  make progress  in the  limit of
small co-dimension $D\to 0$. Note that for $D=0$, i.e.\ for the global
persistence,  Eq.\ (\ref{auto2})  becomes  a pure  exponential  $A(T)=
\exp[-(d-2)T/4]$  for   all  $T$,  indicating  that   the  process  is
Markovian\cite{MBCS}. One then finds $P(T)\sim \exp(-\theta_0 T)$ with
$\theta_0=(d-2)/4$\cite{MBCS}.   For $D$  nonzero but  small,  one can
expand the  correlator in Eq.\ (\ref{auto2}) around the Markov process
($D=0$)  and then  use a  perturbation theory  result\cite{perturb} to
calculate $\theta$ to first order  in $D$.  We  get $\theta= \theta_0
+ D \theta_0^2 I_d/\pi +O(D^2)$ for  all $2<d\leq 4$, where $I_d$ is a
complicated $d$-dependent  integral. For  special values of  $d$, this
integral  simplifies\cite{unpub}.  For  example,  for $d=4$,  we  get,
$\theta= 1/2 + (2\sqrt 2-1)D/4  +O(D^2)$ and for $d=3$, $\theta= 1/4 +
0.183615\dots D +O(D^2)$.

For $D>2$, it is evident from Eq.\ (\ref{auto1}) that one needs to keep
a nonzero cut-off  $a$ in order that the integral  does not diverge at
$t_1=t_2$. In this case, for large $t_1$, the dominant contribution to
the integral comes from the regime $t'\to t_1$. It is easy to see that
in  the limit when  $t_1, t_2$  are both  large with  their difference
$|t_1-t_2|$ fixed, the autocorrelator in Eq.\ (\ref{auto1}) reduces to
a stationary  one, $C(t_1,t_2) \approx B_1(|t_1-t_2|+a^2)^{-(D-2)/2}$,
where  $B_1$ is  an unimportant  constant.   One can  then invoke  the
Newell-Rosenblatt  theorem\cite{NR}  once again  to  conclude that  for
large $t$,  the persistence $P(t)\sim \exp(-\kappa_1 t)$  if $D>4$ and
$\exp(-\kappa_2   t^{(D-2)/2}\ln  t)  \leq   P(t)\leq  \exp(-\kappa_3
t^{(D-2)/2})$ for $2<D<4$, where the  $\kappa_i$'s    are constants. 
Combining these results we thus find that the $n\to \infty$
results for  the persistence of manifold magnetization  in $2<d\leq 4$
are also compatible with our general results in Eq.\ (\ref{asymp1}) on
noting that $\zeta=(D-2)/2$ since  $\eta=0$ and $z=2$ within the large
$n$ limit.

Taking  hints from the  two solvable  cases above  we now  construct a
general  scaling  theory  valid  for  all $d\geq  2$.  The  two-point
correlation function  of the order  parameter, at the  critical point,
has the generic scaling form, $\langle \phi({\bf 0}, t_1)\phi({\bf x},
t_2)\rangle  \sim  x^{-(d-2+\eta)}F(xt_1^{-1/z},  t_2/t_1)$ for  large
distance $x$  and large times $t_1$,  $t_2$, where $\eta$  and $z$ are
the  standard critical  exponents  defined in  the introduction.  This
indicates that  in the Fourier space, $\langle  {\tilde \phi}({\bf K},
t_1){\tilde   \phi}(-{\bf   K},   t_2)\rangle  \sim   K^{-(2-\eta)}G(K
t_1^{1/z},  t_2/t_1)$, where  ${\bf  K}$ is  a $d$-dimensional  vector
conjugate to  $\bf x$.  The manifold magnetization  $\psi$ is obtained
by summing the  order parameter $\phi$ over $d'$  directions.  This is
equivalent   to   putting    $K_i=0$   along   the   $i=1,\dots,   d'$
directions. One  then obtains  the scaling behavior  of the  two point
correlator of the  manifold magnetization, $\langle {\tilde \psi}(\vec
k,      t_1){\tilde      \psi}(-\vec      k,     t_2)\rangle      \sim
k^{-(2-\eta)}g(kt_1^{1/z}, t_2/t_1)$, where $\vec k$ is now a $D=d-d'$
dimensional  vector. For  example, within  the mean-field  theory, the
scaling function  $g(x,y)=\exp[-x^2|1-y|]-\exp[-x^2(1+y)]$, as evident
from Eq.\ (\ref{corrk1}) after dropping the $\Delta$ dependent term at
late times.  The autocorrelation function $C(t_1,t_2)=\langle\psi(\vec
r, t_1)\psi(\vec r, t_2)\rangle$  is then obtained by integrating over
$\vec k$,
\begin{equation}
C(t_1,t_2)=\int {{d^Dk}\over  {k^{2-\eta}}}\,g(kt_1^{1/z}, t_2/t_1)
e^{-k^2a^2},
\label{auto3}
\end{equation}
where $a$ is the soft ultraviolet cut-off as before.

Consider  first the  case  when  $D-2+\eta<0$. One  can  then set  the
cut-off $a=0$ (since the  integral in Eq.\ (\ref{auto3}) is convergent
at    the   upper    limit),   and    one    obtains   $C(t_1,t_2)\sim
t_1^{-(D-2+\eta)/z}f(t_2/t_1)$ in the limit $t_1, t_2 \to \infty$ with
$t_2/t_1$  arbitrary.  Note that  the function  $f(x)\sim
x^{-\lambda_c/z}$   for   large    $x$   such   that   $C(t_1,t_2)\sim
t_2^{-\lambda_c/z}$ for $t_2 \gg t_1$ where $\lambda_c$ is the standard
autocorrelation   exponent\cite{Huse,JSS,Bray}.    This  nonstationary
Gaussian correlator  can then be  reduced, as before, to  a stationary
one for the  normalized variable $X=\psi/\sqrt{\langle \psi^2\rangle}$
in  the logarithmic  time,  $T=\ln t$  and  one gets,  $A(T)=\langle
X(0)X(T)\rangle  =  \exp[(D-2+\eta)T/2z]f(e^T)/f(1)$.  Since  $A(T)\sim
\exp\left[-{(\lambda_c  -(D-2+\eta)/2)}T/z\right]$  for  large  $T$,  it
follows, as  before, that  the persistence $P(T)\sim  \exp(-\theta T)$
for large $T$.  This means that the persistence decays  as a power law
in  the original  time  variable $t=e^T$,  $P(t)\sim t^{-\theta}$  for
large $t$.

In the complementary case $D-2+\eta>0$, the integral in Eq.\ (\ref{auto3}) 
is, for $t_1=t_2$, divergent near the upper limit without the cut-off. 
Hence one needs to  keep $a$ nonzero and then the appropriate scaling 
limit is obtained by taking $t_1, t_2$ both large keeping their 
difference $|t_1-t_2|$ fixed but arbitrary. Then one  can replace the
scaling  function  $g(kt_1^{1/z}, t_2/t_1)$  in  Eq.\ (\ref{auto3}) by
another  function $g_1(k|t_1-t_2|^{1/z})$  of a  single scaling 
variable,  as  in the  two  previous  solvable  cases. Performing  the
integral, one then finds $C(t_1,t_2)\sim |t_1-t_2|^{-(D-2+\eta)/z}$    
for $|t_1-t_2| \gg a^2$. This correlator  is stationary  and decays as 
a  power law.   Invoking the Newell-Rosenblatt  theorem  once  more,  
we find  that  $P(t)$  decays exponentially for $(D-2+\eta)/z >1$  and 
as a stretched exponential for $0 <  (D-2+\eta)/z <1$.  Combining this 
with the  result for  $D-2+\eta <0$ outlined in the previous paragraph, 
gives our general result in Eq.\  (\ref{asymp1}) on   defining   $\zeta=
(D-2+\eta)/z$.
\begin{figure}
  \narrowtext\centerline{\epsfxsize\columnwidth \epsfbox{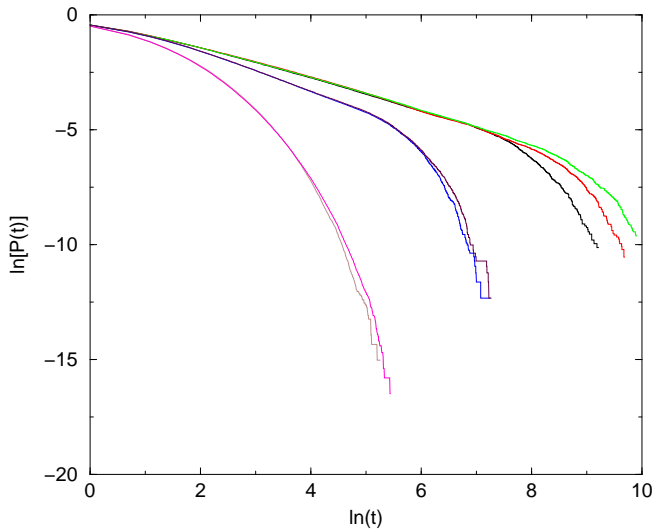}}
\caption{The three outer curves  represent the persistence of the line
magnetization in $d=2$ critical Ising  model as a function of time for
system sizes  $L=63$, $95$  and $127$.  The  two curves in  the middle
represent the persistence of the plane magnetization in $d=3$ critical
Ising  model for system  sizes $L=15$  and $31$.  For the  same system
sizes, the two  leftmost curves represent the persistence  of the line
magnetization in $d=3$.}
\end{figure}

\vspace{-0.2cm}

To test the  analytical  predictions  (i) and  (ii),  we have  done
preliminary Monte  Carlo simulations for  $d=2$ and $d=3$  Ising model
for small size lattices with periodic boundary conditions. The results
are summarized  in Fig.\ 1. The systems  were evolved  using heat-bath
Monte  Carlo  dynamics at  their  bulk  critical couplings:  $K_c=[\ln
(1+{\sqrt  2})]/2$ in $d=2$ and $K_c \approx 0.221656$ in  $d=3$.  For
$d=2$,  the data  for the  persistence of  the line  magnetization for
system  sizes  $L=63$, $95$  and  $127$ (in  each  case  the data  was
averaged  over  $1000$ samples)  shows  a  power  law decay  $P(t)\sim
t^{-\theta}$ for $t \ll L^{z}$ and crosses over to a  faster decay for
$t\gg L^z$. A  fit to the linear  part of the data on  the log-log plot
gives an estimate $\theta\approx  0.72$. Similarly the persistence for
the plane  magnetization in $d=3$  shows a power law  decay, $P(t)\sim
t^{-\theta}$ with $\theta \approx  0.88$, estimated from small lattice
sizes $L=15$ and $31$. The estimates of these exponents are only rough
and may  shift a little with  larger lattices. The  persistence of the
line magnetization  in $d=3$,  in contrast, has  a much  faster decay.
Note that our theory  predicts a decay, $P(t)\sim \exp(-a_1t^{\zeta})$
where  $\zeta\approx  0.016$.  Such  a small  stretching  exponent  is
difficult to determine from the  small size lattices due to the strong
finite-size effects at late times.  All we can say is that the present
data for the line magnetization in $d=3$ is consistent with a decay of
persistence  that  is faster  than  a power  law  but  slower than  an
exponential. More  extensive simulations and  a somewhat sophisticated
finite  size scaling  analysis are  required to  pin down  the precise
value of the stretching exponent\cite{unpub}.

In summary,  we have shown that  the persistence of  submanifolds in a
critical  system decays  with time  in a  manner that  depends  on the
dimensions $d$ and $d'$ of  the system and manifold respectively.  The
crossover  between  the  power-law  decay observed  for  the  `global'
persistence  ($d'=d$)  and  the   exponential  decay  of  the  `local'
persistence ($d'=0$),  occurs via an intermediate  regime of stretched
exponential decay with stretched exponent $\zeta=(d-d'-2+\eta)/z$, for
$0 \le  \zeta \le 1$. This  latter behavior is predicted  for the line
magnetization of the $d=3$ critical Ising model. Numerical simulations
are consistent with the analytical predictions.

\end{multicols}

\end{document}